\documentclass[fleqn,twoside]{article}
\usepackage{espcrc2}

\usepackage{graphicx}
\usepackage[figuresright]{rotating}

\newcommand{\AmS}{{\protect\the\textfont2
  A\kern-.1667em\lower.5ex\hbox{M}\kern-.125emS}}

\title{The MIDAS experiment: A prototype for the microwave emission of Ultra-High Energy Cosmic Rays}

\author{M. Monasor\address[MCSD]{Department of Astronomy \& Astrophysics and Kavli Institute for Cosmological Physics\\
        University of Chicago ,
        5640 S Ellis Ave, 60637 Chicago, IL, U.S.A.}%
        \thanks{The author thanks the Consejeria Educaci{\'o}n y Ciencia de Castilla-La Mancha and FSE for a postdoctoral fellowship.},
        I.~Alekotte\address[Argentina]{Centro At\'{o}mico Bariloche and Instituto Balseiro, San Carlos de Bariloche, Argentina},
        J.~Alvarez-Mu\~{n}iz\address[Santiago]{Universidad de Santiago de Compostela, Spain},
        A. Berlin\addressmark[MCSD],
        X.Bertou\addressmark[Argentina],
        M. Bodgan\addressmark[MCSD],
        M. Bohacova\addressmark[MCSD],
        C. Bonifazi\address[Brazil]{Universidade Federal do Rio de Janeiro, Instituto de F\'{\i}sica, Rio de Janeiro, Brazil},
        W. Carvalho\addressmark[Santiago],
        J.R.T.~de Mello Neto\addressmark[Brazil],
        J.F. Genat\addressmark[MCSD],
        P. Facal San Luis\addressmark[MCSD],
        E. Mills\addressmark[MCSD],
        B. Rouille d'Orfeuil\addressmark[MCSD],
        S. Wayne\addressmark[MCSD],
        L.C. Reyes\addressmark[MCSD],
        E.M.~Santos\addressmark[Brazil],
        P. Privitera\addressmark[MCSD],
        C. Williams\addressmark[MCSD]
        and
        E. Zas\addressmark[Santiago]
        }
       
\begin{document}

\begin{abstract}
Recent measurements suggest that extensive air showers initiated by ultra-high energy cosmic rays (UHECR) emit signals in the microwave band of the electromagnetic spectrum caused by the collisions of the free-electrons with the atmospheric neutral molecules in the plasma produced by the passage of the shower. Such emission is isotropic and could allow the detection of air showers with 100\% duty cycle and a calorimetric-like energy measurement, a significant improvement over current detection techniques. We have built MIDAS (MIcrowave Detection of Air Showers), a prototype of microwave detector, which consists of a 4.5 m diameter antenna with a cluster of 53 feed-horns in the 4 GHz range. 
The details of the prototype and first results will be presented.
\vspace{1pc}
\end{abstract}

\maketitle

\section{INTRODUCTION}

One hundred years after their discovery the properties of UHECR, the most energetic particles in the universe, remain still uncertain. Their sources, nature and acceleration and propagation mechanisms are nowadays some of the major open questions in astrophysics. A great effort has being carried out with experiments such as HiRes~\cite{bib:hires} and AGASA~\cite{bib:agasa} and at present the Pierre Auger Observatory~\cite{bib:auger} to acumulate statistics in a energy region (well above 1 EeV) where the rate of events is extremely low due to the suppression of the flux, recently observed and confirmed~\cite{bib:hires_gzk,bib:auger_gzk}, when UHECR interact with the cosmic microwave radiation.   

When an UHECR interacts with an atmospheric molecule it developes an Extensive Air Shower (EAS). The particles in the shower deposit part of the energy carried by the primary in the atmosphere through ionization, resulting in a free electron plasma.  Microwave emission due to the molecular bremsstrahlung of free-electron collisions with the neutral molecules of the atmosphere has been measured in a particle accelerator and could be used to detect air showers \cite{Gorham08}.  A cosmic ray detector based on this technique would provide calorimetric measurement of the primary energy and a direct measurement of the maximum development~\footnote{An observable directly related to the mass of the primary.} $X_{max}$, main advantages of the fluorescence detectors used at present in cosmic ray experiments. This technique has also a nearly 100\% duty cycle, as surface detectors sampling the shower at ground level, and very little atmospheric attenuation. 

The MIDAS experiment, a prototype detector to characterize this microwave emission, has been recently commissioned at the University of Chicago. 
 Design, specifications of the electronic components, a description of the implemented trigger logic, and preliminary test results are discussed below.

\section{THE MIDAS PROTOTYPE}

The main goal of the MIDAS experiment is to detect and characterize the microwave emission from EAS. For this purpose, we have built a wide field of view telescope with a camera of 53 channels of 0.8 GHz bandwidth in the 4 GHz microwave band. The center of the camera is located at the focus of a 4.5 m diameter parabolic antenna. Each channel is equipped with a C-band commercial satellite television receiver, i.e. a combination of feed horn, low noise amplifier, and frequency down converter. The signal from each of these receivers is then transformed to a voltage level using a fast power detector with logarithmic response to the input RF power. The DC output signal (ranging from 0.5-2.1 V) is continuously digitized in custom 14-bit 20 MHz Analog-to-Digital Converters (ADC) modules. Digital samples are processed locally with on-board Field Programmable Gate Arrays (FPGAs), where a trigger logic allows discrimination of events. A trigger master board is used to synchronize all ADC boards, and to generate a global trigger. It provides a system clock for simultaneous sampling and receives the local trigger from each ADC board via front LVDS-panel connections. After a trigger, a 100 microsecond time stream of data (25 $\mu$s corresponding to data before the trigger) is packed and buffered for readout via a VME back panel and stored for later analysis. A VME GPS module (Hytec 2092) is also used for precise timing information.

 

\subsection{Trigger and DAQ systems}

The 53 receiver channels are serviced by 4 identical ADC boards. 
On-board FPGAs allow local data analysis, based on a trigger level system developed to select candidate events. The trigger system is based on the same principles as the one used for the fluorescence detector of the Pierre Auger Observatory~\cite{bib:auger_report}. 

\begin{figure}[htb]
\vspace{-9pt}
\centering{\includegraphics[width=\columnwidth]{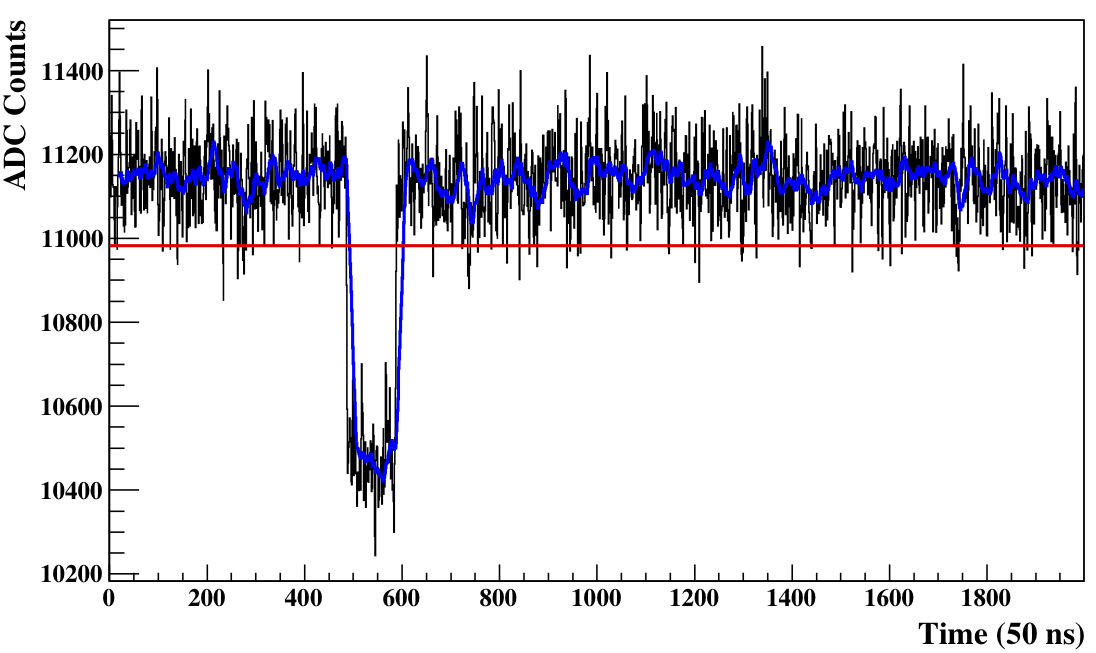}}
\vspace{-30pt}
\caption{Response of one of the channels to a pulse generated by a transmiting antenna. The ADC trace (black line), running sum (blue line) and threshold (red line) are presented.}
\label{fig:flt}
\end{figure}

A first level trigger (FLT) at the channel level, has been designed to identify peaks in the FADC trace. The FLT information of all channels is then sent to the trigger master board to perform a second level trigger (SLT) at the level of the whole camera. If the SLT condition is satisfied, the system trigger pulse is send to the 4 ADC boards for the readout. These first and second level trigger conditions are implemented as follows:

FLT: A channel is marked as triggered if the running sum of 20 consecutive bins exceeds an adjustable threshold. This status is extended for 10 $\mu$s to allow coincidences between channels in subsequent trigger requirements. The threshold is continuously regulated to compensate the changing background conditions and to ensure a constant trigger rate per channel of 100 Hz. The procedure is illustrated in figure \ref{fig:flt}. 

\vspace{-10pt}

SLT: The FLT information is sent to the Master Trigger board where another on-board FPGA is used to evaluate trigger conditions at the level of the whole camera. The SLT block is designed to search for 4-fold patterns of FLT channels compatible with an UHECR shower track. A SLT rate of much less than 1 Hz is expected from random coincidences.

High level trigger conditions have been also developed to deal with the different background conditions. For example, the threshold regulation and data acquisition are inhibited when SLT rate is higher than a certain limit. 

\begin{figure}[htb]
\vspace*{-20pt}
\centering{\includegraphics[width=\columnwidth]{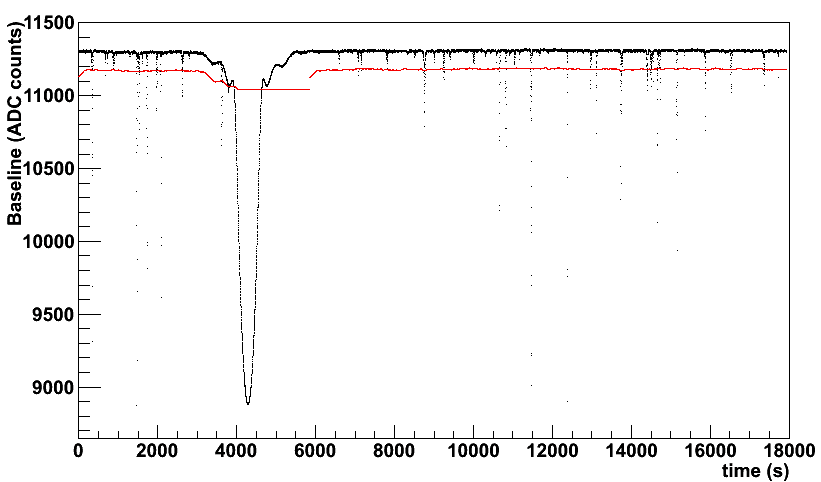}}
\vspace*{-25pt}
\caption{Average baseline (black line) of the central channel of the camera during 5 hours. The threshold (red line) follows the baseline structure except when the Sun enters in the field of view due to the high trigger rate.}
\label{fig:sun}
\end{figure}

The DAQ system is continuosly monitoring the data conditions by recording each second an average baseline (over 10 ms) for each channel, the instantaneous threshold and FLT rate. Figure \ref{fig:sun} shows the temporal evolution of the baseline for the central channel of the camera with the adjustable threshold for the same period. As it can be observed, the threshold reproduces accurately the baseline profile. The large peak shown in the plot is originated by the Sun passing through the field of view of the feed. The fast peaks in the figure are noise bursts, possible induced by aircrafts, which reduce significantly the duty cycle of the experiment. 
 
\vspace{-15pt}

\subsection{Calibration}
An absolute calibration of the antenna is provided by the Sun, whose
emission in the frequency of interest is measured every day by several
radiotelescopes around the world. Figure \ref{fig:sun} shows the signal in ADC counts generated by the Sun crossing the field of view of the central channel of the camera. The system temperature can then be derived from the difference in ADC counts between the peak induced by the Sun and the baseline. A patch antenna located at the center of the 4.5 m dish is used to generate RF signals of known intensity, thus performing a relative calibration of the individual feed channels. Also, it allows the study of the relative timing, which was found to be well below the 50 ns sampling.


\subsection{Data taking}

Up to now, the telescope has been taking data in stable conditions during 40 days with an average event rate of 0.5 Hz. A prelimary analysis suggests that data are dominated by noise, both the expected uncorrelated random noise and correlated noise due to fast transient events of unknown origin. All channels present a good long term stability with baseline fluctuations smaller than 1 dB.
Physics event search is still in progress. Events simulations have been also carried out including the measured values of the detector calibration, the noise level and the microwave emission scaled from the laboratory measurements. 




\section{CONCLUSIONS}
The MIDAS experiment has been assembled and commissioned between October 2009 and February 2010 at the University of Chicago. 
The first period of data taking has been focused on the understanding of the detector and of the noise conditions. The sensitivity of the prototype was found to be adequate for detection of UHECR. Some improvements are being implemented to increase the duty cycle to close to 100\%. In order to crosscheck candidate events with already established detection techniques, we foresee the operation of MIDAS at the southern Auger Observatory.

\end{document}